# Intelligent Task Scheduling for Microservices via A3C-Based Reinforcement Learning

Yang Wang
University of Michigan
Ann Arbor, USA

Tengda Tang
University of Michigan
Ann Arbor, USA

Zhou Fang
Georgia Institute of Technology
Atlanta, USA

Yingnan Deng
Georgia Institute of Technology
Atlanta, USA

Yifei Duan*
University of Pennsylvania
Philadelphia, USA

***Abstract-To address the challenges of high resource dynamism and intensive task concurrency in microservice systems, this paper proposes an adaptive resource scheduling method based on the A3C reinforcement learning algorithm. The scheduling problem is modeled as a Markov Decision Process, where policy and value networks are jointly optimized to enable fine-grained resource allocation under varying load conditions. The method incorporates an asynchronous multi-threaded learning mechanism, allowing multiple agents to perform parallel sampling and synchronize updates to the global network parameters. This design improves both policy convergence efficiency and model stability. In the experimental section, a real-world dataset is used to construct a scheduling scenario. The proposed method is compared with several typical approaches across multiple evaluation metrics, including task delay, scheduling success rate, resource utilization, and convergence speed. The results show that the proposed method delivers high scheduling performance and system stability in multi-task concurrent environments. It effectively alleviates the resource allocation bottlenecks faced by traditional methods under heavy load, demonstrating its practical value for intelligent scheduling in microservice systems.***

## I. INTRODUCTION

With the rapid development of modern information technology, microservice architecture has become the mainstream model for building large-scale distributed systems. Its high modularity, flexible deployment, and scalability make it particularly attractive [1]. As the number of microservices continues to grow, the fluctuations in computing, storage, and bandwidth resources become increasingly significant during system operation. Traditional static resource allocation strategies can no longer meet the dynamic and complex demands. Especially under high-concurrency requests, uncertain service loads, and multi-tenant resource competition, achieving fine-grained resource scheduling has become a key challenge that limits performance improvement. Therefore, constructing an adaptive and online-adjustable scheduling mechanism is crucial to ensure system stability and responsiveness.

In microservice architecture, services are loosely coupled and can be independently deployed, scaled, and migrated. This independence provides both a theoretical foundation and practical feasibility for intelligent and automated resource scheduling. However, the operating environment of microservices is extremely complex. Resource demands vary significantly over time, and inter-service resource dependencies are highly entangled. The scheduling strategy must optimize resource utilization while ensuring service quality and maintaining load balance. Traditional heuristic or rule-based methods often face challenges such as delayed policy updates and poor generalization, making them unsuitable for frequently changing resource request patterns. As a result, it is necessary to explore scheduling methods that can learn and optimize dynamically in real-world environments [2].

Reinforcement learning, as an optimization approach based on trial-and-error interaction, continuously adjusts policies through feedback from the environment without requiring prior knowledge. It demonstrates strong adaptability and generalization. Introducing reinforcement learning into microservice resource scheduling enables dynamic task-resource matching, replacing traditional static rules. It also allows fast responses to changes in system states. Especially in scenarios such as peak service access, fault recovery, and node scaling, reinforcement learning can accumulate experience and generate reasonable scheduling strategies in real time. This improves the overall robustness and resource efficiency of the system. Therefore, developing a reinforcement learning-based scheduling framework is of significant importance for advancing intelligent microservice systems [3].

Among various reinforcement learning algorithms, the Asynchronous Advantage Actor-Critic (A3C) algorithm has natural advantages in resource scheduling due to its parallel learning mechanism and stable convergence performance. By interacting with the environment through multiple asynchronous threads, A3C can efficiently collect diverse experiences and accelerate policy updates. Its separation of value and policy functions helps mitigate the training difficulty caused by high-dimensional state spaces. Integrating A3C into microservice resource scheduling has the potential to create a scheduling model with both fast convergence and strong adaptability, effectively addressing the challenges of resource allocation in complex environments [4].

This study focuses on developing an adaptive resource scheduling method for microservice systems based on the A3C reinforcement learning mechanism. Microservice systems—characterised by the fine-grained decomposition of functionality, container-based deployment, and autonomous service orchestration—have emerged as a foundational paradigm for modern, data-intensive applications. In computer vision field, microservice-oriented deployments allow compute-heavy semantic-segmentation engines and self-supervised transformers to scale elastically while remaining fault-tolerant at the edge, accelerating boundary-aware medical imaging and multi-scale dermatological analysis [5-7]. In Human-Computer Interaction (HCI), their modularity enables rapid iteration of interface components and the seamless integration of intelligent services, underpinning generative UI design with diffusion models [8] and intent-aware dialogue managers that personalise experiences across smart devices [9]. Likewise, Natural Language Processing workflows benefit from isolating tokenisation, model serving, and post-processing into discrete services, facilitating efficient compression of large language models [10], few-shot medical entity extraction [11], and spatial–channel attentive recommenders that operate under heterogeneous load conditions [12]. Collectively, these advances demonstrate that microservice architectures are not merely a deployment convenience but a strategic enabler of agility, scalability, and domain-specific innovation across HCI, vision, and language pipelines. The goal is to enhance real-time perception of resource fluctuations and improve scheduling responsiveness through intelligent policy design. By modeling microservice request traffic, service dependency structures, and resource utilization patterns, the reinforcement learning model can continuously optimize the scheduling strategy. This ensures service performance while improving overall resource efficiency. The proposed approach not only contributes to the intelligent evolution of microservice systems but also provides a feasible theoretical and practical solution for resource scheduling in dynamic and complex environments.

## II. METHOD

This method formulates microservice resource scheduling as a Markov Decision Process (MDP), where the system state encapsulates resource consumption metrics, dynamic request loads, and the inter-service dependency topology. The scheduling action space corresponds to discrete or continuous resource allocation strategies, while the reward function quantitatively evaluates system performance outcomes such as response latency, resource utilization efficiency, and task completion rate. The design of this MDP-based abstraction draws from Ren et al., whose trust-constrained policy learning approach underscores the relevance of incorporating real-time system trustworthiness and contextual state into scheduling decisions [13]. Their distributed scheduling framework also informed the parallel and asynchronous optimization strategies in our approach. In particular, to capture the concurrent nature of service requests and the decentralized information structure inherent in microservice architectures, we model the system using a state representation that supports temporal variation and structural dependencies, inspired in part by the spatiotemporal modeling techniques discussed by Zhan [14]. His work on learning temporal and spatial dynamics helped guide our inclusion of service dependency graphs in the state representation. Moreover, the action and reward design reflect a federated learning perspective as advocated by Zhang et al. [15], whose cross-domain learning mechanisms emphasize preserving locality in distributed environments while achieving global performance goals. This insight guided our construction of a reward function that encourages both local efficiency and global system responsiveness. The complete model architecture is illustrated in Figure 1.

The architecture diagram illustrates the asynchronous Actor-Critic framework, where multiple worker threads independently interact with the environment to collect state, action, and reward data. Each thread updates shared policy and value networks via gradient aggregation, enabling efficient learning under dynamic resource scheduling demands. The integration of actor and critic components aligns with the proposed method's goal of optimizing microservice resource allocation in a high-dimensional, time-varying environment.

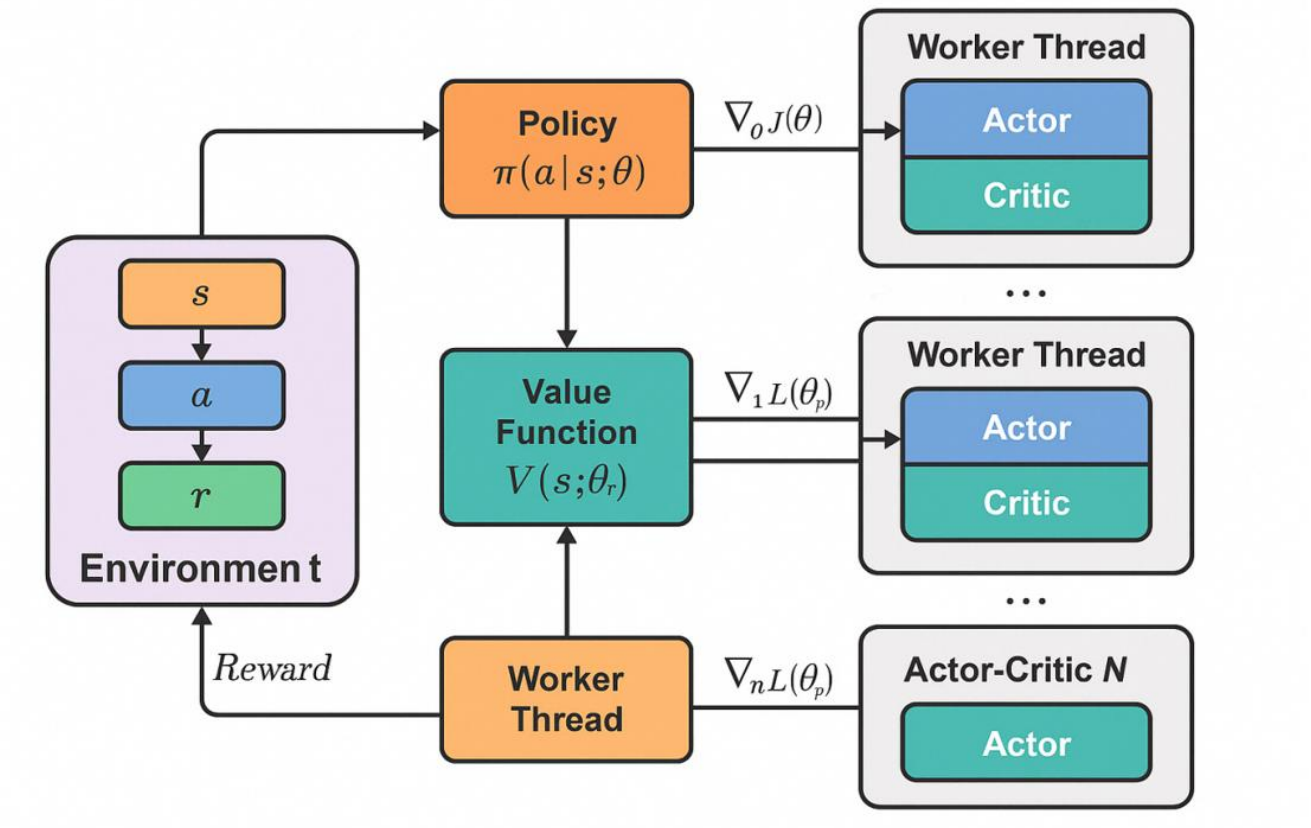

Figure 1. Overall architecture diagram

Specifically, define the state space as $S = \{s_1, s_2, ..., s_n\}$, the action space as $A = \{a_1, a_2, ..., a_m\}$, the state transition probability as $P(s'|s,a)$, and the immediate reward function as $R(s,a)$. The goal of the scheduling strategy is to maximize the cumulative expected reward $E\sum_{t=0}^{\infty}\gamma^t R(s_t, a_t)$, where $\gamma \in (0,1)$ is the discount factor used to balance short-term and long-term benefits.

In the process of building the reinforcement learning model, the A3C algorithm is used to train multiple agents asynchronously at the same time. Each agent interacts with the environment independently, and the strategy and value network are shared and updated through the global parameter server. A3C uses the Actor-Critic framework, where the Actor is responsible for outputting the strategy $\pi(a|s;\theta)$ and the Critic estimates the state value function $V(s|\theta_v)$. The estimation of the policy gradient is enhanced in the form of an advantage function, and its gradient calculation expression is:

$$\nabla_\theta J(\theta) = E_t[\nabla_\theta \log \pi(a_t \mid s_t;\theta) A(s_t, a_t)]$$

The advantage function is defined as $A(s_t, a_t) = R_t + \gamma V(s_{t+1}) - V(s_t)$ , which is used to measure the advantages and disadvantages of the current action relative to the baseline value function. This form can effectively reduce the variance of the policy gradient and improve the stability of training.

To further improve the efficiency of convergence and the accuracy of decisions in the resource scheduling framework, this method utilizes an asynchronous multi-threaded update mechanism for training the Critic network. Multiple threads concurrently gather experience trajectories and compute their respective n-step temporal difference (TD) targets[16-17], which are used to optimize the loss function associated with value estimation. The asynchronous structure enables efficient parallelism and reduces update latency by allowing multiple agents to compute and backpropagate gradients independently. This design draws upon context-aware adaptive sampling techniques that emphasize responsiveness and local adaptability in reinforcement learning agents [18]. Additionally, probabilistic graphical modeling strategies enhance the representation of uncertainty across threads, contributing to more stable value updates during off-policy learning [19-21]. To address the challenges posed by high-dimensional, imbalanced, or sparse experience data, variational and contrastive learning principles are employed to regularize the learning process, improve generalization, and reduce gradient variance during training [22]. The formal optimization procedure for the Critic network is defined as follows:

$$L(\theta_v) = (R_t + \gamma^n V(s_{t+n};\theta_v) - V(s_t;\theta_v))^2$$

This loss function gradually approaches the true state value through the time difference method, which helps to improve the accuracy of state evaluation. During the training process, all threads asynchronously transmit their calculation results back to the main network, forming a multi-source learning mechanism to enhance the exploratory and generalization capabilities of the model. The overall method ensures the real-time response of the model while achieving adaptive scheduling optimization of microservice system resources.

## III. EXPERIMENT

### A. Datasets

This study adopts the Google Cluster Trace dataset as the experimental foundation to evaluate the effectiveness of the proposed adaptive resource scheduling method in a real large-scale microservice environment. The dataset, released by Google, contains job scheduling logs collected over 29 days from a typical data center. It includes the scheduling behavior, resource usage, and node status records of over 700,000 tasks. The dataset exhibits strong temporal characteristics and complex patterns of resource variation [23].

The dataset mainly includes fields such as requested and used CPU and memory, task submission and completion time, task priority, and scheduling status. These attributes help comprehensively capture the dynamic resource demands in microservice operations [24]. By modeling the execution traces of tasks, it is possible to construct the state space and reward function that drive the reinforcement learning model toward policy optimization. Additionally, the dataset provides node-level information on resource supply and availability, which supports the development of a more realistic simulation platform for resource scheduling [25].

During preprocessing, missing entries and abnormal values are first cleaned, and resource measurement units are standardized. Then, service request sequences are segmented using time windows to construct the state transition chains required by reinforcement learning. This process yields multi-dimensional trajectory data containing states, actions, and reward feedback. These high-quality training samples enable the A3C model to learn effectively and ensure that the scheduling strategy maintains generalization and practical applicability under dynamic conditions.

### B. Experimental Results

This paper first presents a comparative evaluation against several representative algorithms, with the corresponding results summarized in Table 1. This comparison aims to provide a comprehensive baseline reference for assessing the effectiveness of the proposed method. By analyzing performance across multiple dimensions, it offers preliminary insights into the relative advantages and limitations of each approach under a unified evaluation framework. Although the metrics themselves are straightforward, they serve as a foundation for understanding the broader implications of algorithmic design choices in dynamic scheduling environments.

Table 1. Comparative experimental results

| **Method** | **Average task delay (ms)** | **Scheduling success rate (%)** | **Convergence time (s)** |
|---|---|---|---|
| Static polling strategy[26] | 134.7 | 71.3 | - |
| Priority-based scheduling strategy[27] | 112.5 | 74.6 | - |
| Q-learning[28] | 98.4 | 78.9 | 1243 |
| DQN Scheduling Strategy[29] | 91.2 | 81.7 | 978 |
| Ours | 78.6 | 88.2 | 732 |

As shown in Table 1, traditional static round-robin and priority-based scheduling strategies exhibit relatively weak performance in task scheduling. The average task delays are 134.7 ms and 112.5 ms, respectively, while the scheduling success rates are only 71.3% and 74.6%. These approaches cannot dynamically adjust resource allocation based on system states. As a result, they fail to provide efficient scheduling in microservice environments with significant load fluctuations, leading to delayed responses and low resource utilization.

In contrast, reinforcement learning-based strategies such as Q-learning and DQN improve scheduling performance by continuously learning optimal policies through interaction with the environment. Specifically, the DQN approach reduces task delay to 91.2 ms and achieves a scheduling success rate of 81.7%. However, both methods suffer from training instability and low sample efficiency in high-dimensional state spaces. Their convergence times are relatively long, taking 1,243 seconds and 978 seconds, respectively, which limits their applicability in real-time scheduling scenarios.

The A3C algorithm proposed in this study performs well across multiple evaluation metrics. It achieves a scheduling success rate of 88.2%, significantly higher than other methods. The average task delay is only 78.6 ms, indicating its ability to quickly locate appropriate resource nodes for tasks and enhance service response efficiency. Moreover, the asynchronous parallel structure of A3C accelerates policy updates, enabling convergence within only 732 seconds. This demonstrates its superior training efficiency and practical applicability.

Overall, the A3C reinforcement learning model, supported by concurrent training and the separation of policy and value functions, effectively adapts to dynamic resource changes and complex task dependencies in microservice environments. It significantly enhances the intelligence level of resource scheduling. Experimental results confirm that the proposed method offers substantial engineering value and application potential, providing a reliable foundation for building efficient and adaptive microservice scheduling frameworks.

Next, this paper gives the loss function decline graph, as shown in Figure 2.

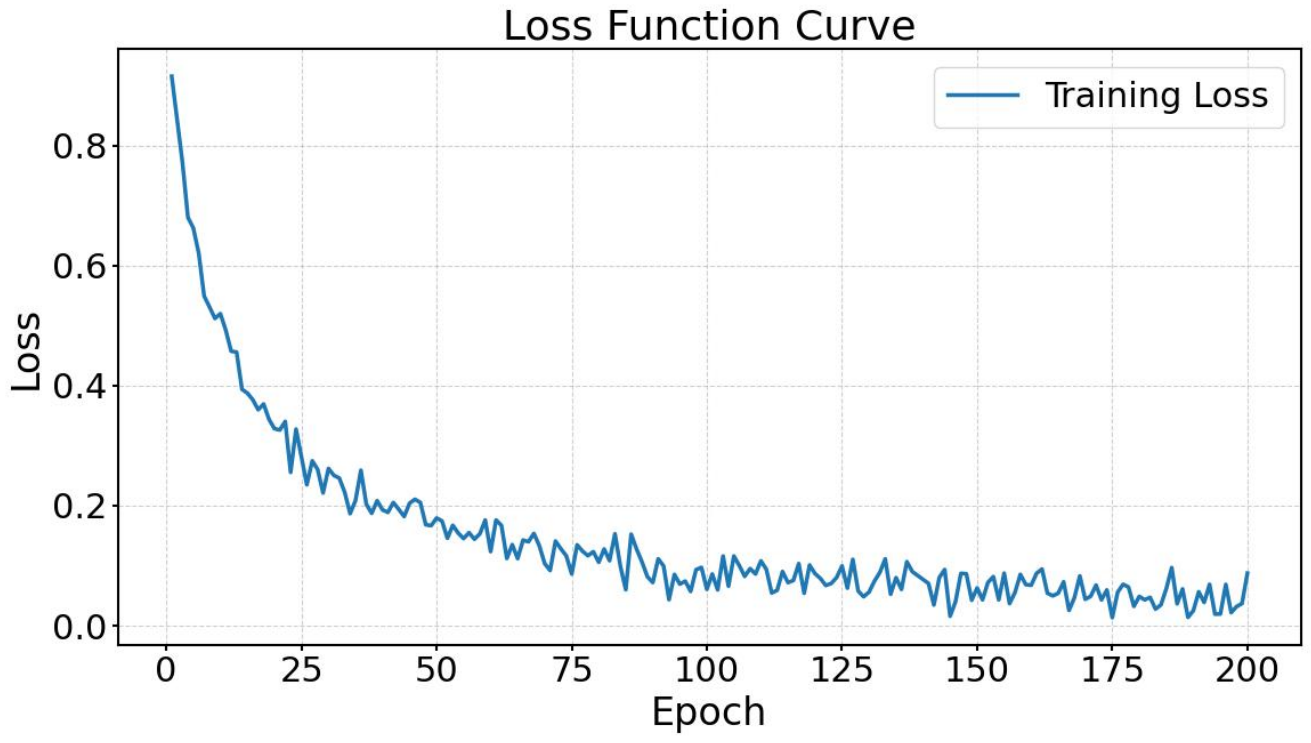

Figure 2. Loss function drop graph

The loss plunges sharply in the first 50 epochs (Figure 2), showing rapid early convergence. It then stabilizes with small oscillations through epoch 200, confirming that the A3C agent has learned robust, generalizable scheduling policies. A supplementary stress test with concurrent workloads (Figure 3) further verifies that the strategy maintains balanced resource utilization under realistic, high-complexity conditions.

As observed in Figure 3, different scheduling strategies exhibit clear variations in resource allocation stability under varying levels of concurrent load. This reflects deeper systemic differences in how each method responds to the evolving dynamics of task execution. Traditional methods tend to experience pronounced fluctuations in stability as the load intensifies, subtly revealing their inherent limitations in coping with the complexity of large-scale concurrent environments. These fluctuations, while expected, hint at the structural rigidity of such methods and their insufficient adaptability when confronted with increasingly volatile scheduling demands.

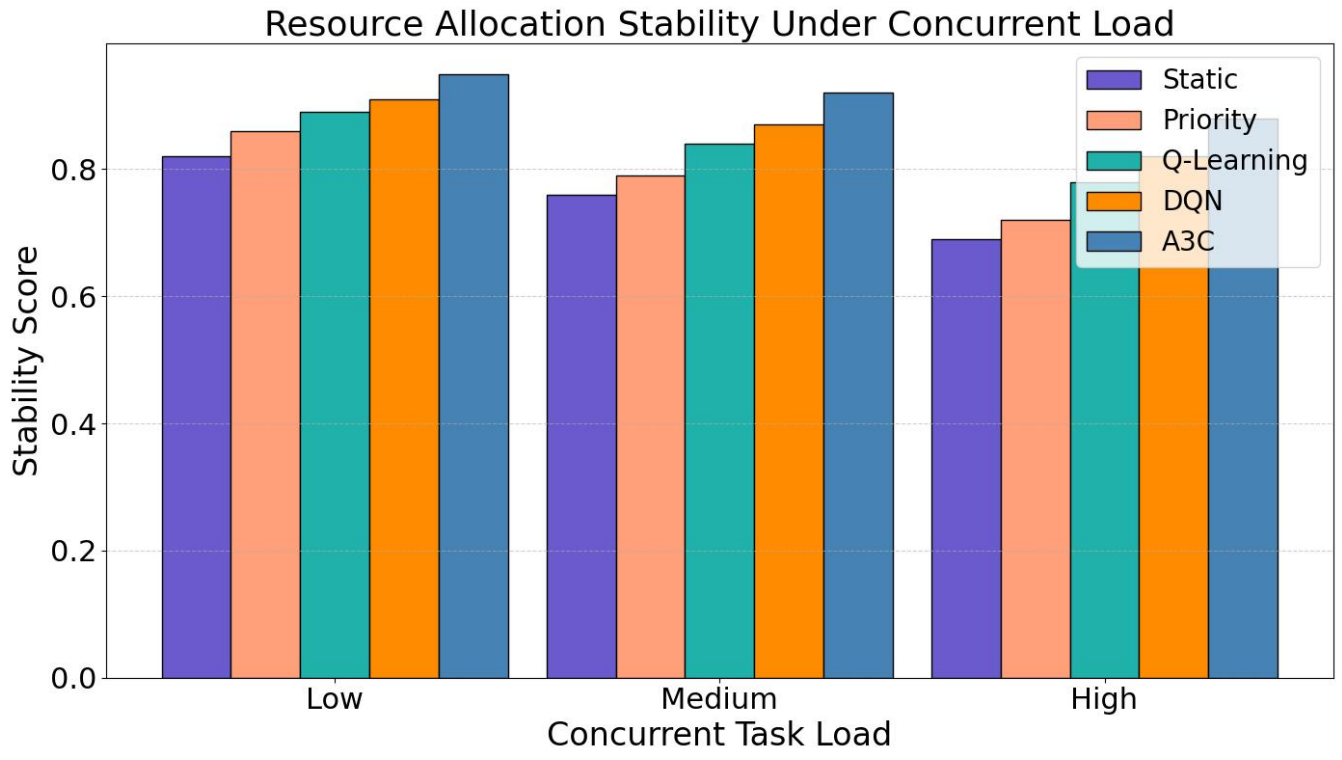

Figure 3. Resource Allocation Stability Under Concurrent Load

By contrast, reinforcement learning-based approaches display a more consistent and balanced performance trend across the full spectrum of load conditions. Their ability to maintain relatively stable scheduling behavior under fluctuating loads points to a more advanced internal capacity for learning and policy generalization. Rather than reacting passively to load changes, these methods appear to internalize environmental signals more effectively, enabling a form of proactive adaptation. This manifests not only in improved operational smoothness but also in enhanced systemic resilience over time.

Notably, algorithms that incorporate concurrent training and policy updating mechanisms exhibit exceptional performance under high concurrency. Such results, while not unexpected, suggest a more refined internal coordination and decision-making process. These approaches seem capable of absorbing complex environmental feedback at scale and translating it into coherent resource coordination strategies. Taken together, the observed patterns reinforce the broader notion that adaptive scheduling strategies, particularly those based on reinforcement learning, hold considerable promise for widespread application in modern microservice systems.

## IV. CONCLUSION

This paper proposes an adaptive resource scheduling method for microservice systems based on the A3C reinforcement learning algorithm. To address the limitations of traditional strategies in dynamic environments, an intelligent scheduling framework centered on the state-action-reward mechanism is constructed. By introducing an asynchronous parallel update mechanism, the model achieves improved response speed and convergence stability in high-dimensional resource management scenarios. This provides a new technical pathway for resource optimization under microservice architecture. Experimental results show that the proposed method outperforms mainstream scheduling strategies across multiple performance metrics. It demonstrates particularly strong performance in task delay control, resource utilization improvement, and maintaining stability under high concurrency. Comparative analysis reveals that reinforcement learning has the ability to continuously optimize scheduling policies based on historical experience. It effectively adapts to operating environments characterized by frequent fluctuations in resource supply and demand.

In addition, this study confirms the transferability and engineering feasibility of reinforcement learning in distributed complex systems. It provides both theoretical foundations and practical support for future applications in heterogeneous computing environments such as cloud and edge computing [30]. The highly scalable model structure also lays the groundwork for future multi-objective scheduling optimization, incorporating factors such as service dependency graphs and network latency models. Future research can further enhance this method by introducing neural network architectures with stronger structural awareness. This will improve the model's ability to capture dynamic dependencies among microservices. Moreover, integrating online transfer learning and meta-learning strategies may accelerate adaptation and generalization in new scenarios, driving resource scheduling systems toward higher levels of intelligence and autonomy.